\documentclass[%
superscriptaddress,
preprint,
 amsmath,
 amssymb,
 aps,
]{revtex4-2}

\usepackage{xr}
\usepackage{graphicx}
\usepackage{dcolumn}
\usepackage{xcolor}
\usepackage{bm}
\newcommand{\ts}{\textsuperscript}
\usepackage[version=4]{mhchem} 
\usepackage{siunitx}
\usepackage{enumitem}

\begin{document}

\preprint{APS/123-QED}

\title{
Imaging ultrafast dynamical diffraction wavefronts in strained Si with coherent X-rays
}

\author{Angel Rodriguez-Fernandez}
\email{angel.rodriguez-fernandez@xfel.eu}
\affiliation{European X-ray Free Electron Laser GmbH, Schenefeld, DE-22986.}

\author{Ana Diaz}%
\affiliation{Paul Scherrer Institute, Forschungsstrasse 111, Villigen PSI, CH-5232}

\author{Anand H. S. Iyer}%
\affiliation{Department of Physics, Chalmers University of Technology, Gothenburg, SE-41296}

\author{Mariana Verezhak}%
\author{Klaus Wakonig}%
\affiliation{Paul Scherrer Institute, Forschungsstrasse 111, Villigen PSI, CH-5232}

\author{Magnus H. Colliander}%
\affiliation{Department of Physics, Chalmers University of Technology, Gothenburg, SE-41296}
 
\author{Dina Carbone}%
\affiliation{ MAX IV Laboratory, Lund University, Lund, SE-22199.}

\date{\today}

\begin{abstract}
Dynamical diffraction effects in thin single crystals produce highly monochromatic parallel X-ray beams with a mutual separation of a few \SI{}{\micro\meter} and a time-delay of a few fs \textemdash the so-called echoes. 
This ultrafast diffraction effect is used at X-ray Free Electron Lasers in self-seeding schemes to improve beam monochromaticity. 
Here, we present a coherent X-ray imaging measurement of echoes from Si crystals and demonstrate that a small surface strain can be used to tune their temporal delay.
These results  represent a first step towards the ambitious goal of strain-tailoring new X-ray optics and, conversely,  open up  the possibility of using ultrafast dynamical diffraction effects to study strain in materials.
\end{abstract}


\maketitle


The emergence of 4\ts{th} generation synchrotron sources based on low-emittance rings  \cite{MAXIV18} and of spatially coherent X-ray Free Electron Lasers (XFELs) \cite{EUXFEL20,LCLS10}, open a new era in the study of high coherence materials, such as perfect crystals, using high coherence probes \cite{Vaclav04}. 
Recent years have seen an increase in the development and use of inverse microscopy techniques based on coherent X-rays \cite{Thibault379,Ulvestad2014, Singer2018, Godard11, Hill18, Hruszkewycz2017, Chamard10}, and other types of speckle analysis \cite{Zhang18, Jacques16}, for the study of crystalline materials , as well as for the characterization of coherent X-ray beams \cite{Schropp2010, Vila2011, Schropp2013,Bjorling2020,Thibault379}.
These methods are based on the assumption of a Fourier transform relation between the sample's electron density and the X-ray scattered field, derived from the so-called kinematical approximation \cite{Zacha1945}.
On the other end, large single crystals are one of the primary optical elements for X-ray experiments.
For these, more complex interactions including multiple diffraction, absorption and refraction effects must be considered, which are well described by the dynamical diffraction theory \cite{Zacha1945, Batter64,Authi01}.
Dynamical diffraction effects have already proven very useful for applications in ultrafast X-ray optics \cite{Amann2012}.
Nevertheless, dynamical diffraction can arise already from samples of modest dimensions ($\sim\SI{500}{\nano\meter}$) and, in conjunction with the use of coherent nanobeams, have already been shown to be not only measurable \cite{Pateras18,Civita18} but also a hindrance to standard inversion algorithms for digital microscopies \cite{Shabalin17} and requiring adaptive solutions \cite{Gorobtsov16}. 

Here, we show that the exceptional coherence properties of the X-ray beam produced at MAX IV Laboratory can be exploited to understand dynamical diffraction effects in crystalline samples and how the wave-front changes in presence of strain, and propose an approach to analyse them. The effect under study, presented theoretically in Ref.~\cite{Shvy12}, consists of the appearance of multiple monochromatic beams in the diffracted and forward directions, that present a transverse displacement of a few microns between each other.
These beams, also known as echoes, are delayed between each other by few femtoseconds depending of the thickness of the crystal and energy of the X-ray beam. 
This ultra-fast effect is used in X-ray optics at XFELs for self-seeding radiation production \cite{Amann2012}. 
We have previously demonstrated the transverse displacement of the echoes in diamond plates \cite{ARF18} and Si wafers \cite{ARF20} using a $\SI{2}{\micro\meter}$ beam focused onto an optically coupled X-ray detector with sub-$\SI{}{\micro\meter}$ resolution. 
In this Letter, we improve the resolution of our previous works and demonstrate that localised surface strain in a Si wafer can produce a spatial modulation of the echoes that translates into the tuning of their time delay.

A nanobeam is used to select regions of a Si sample where a residual strain field was created by nano-indentation. 
The coherence of the X-rays is exploited to measure the finely structured echoes, strongly dependent on the local strain field, with a sub-$\SI{100}{\nano\meter}$  resolution only achievable with inverse microscopy.
The technique used, known as X-ray tele-ptychography \cite{Tsai26}, is capable of measuring the X-ray diffracted field, using as probe a small pinhole placed \emph{after} the sample \footnote{We note that conventional ptychography, in which the sample is scanned with respect to the incoming beam, would not work in presence of dynamical diffraction due the assumption of a factorization of the illumination and the sample transmissivity in conventional ptychography, ref. \cite{Thibault379}}. 
In this way, no assumption is made on the interaction of X-rays with the sample and this approach can  be  extended  to  study  dynamical diffraction effects.
A similar configuration, using a large coherent beam, with the sample in diffraction condition for strain sensitivity, has been recently shown in \cite{Verezhak18, verezhak2020}.
The results presented are corroborated by simulations based on dynamical diffraction theory. 
This work provides a very fine tool to understand the role of dynamical diffraction in the study of ultrafast processes in perfect and strained thin crystals.
With this knowledge new X-ray optics for high coherence sources can be designed when a better control of the temporal dependence of the diffracted signal is needed.
Finally, this work proposes an approach for the study of strained micro-crystals and time-resolved lattice deformation \cite{Crysta_top} in highly ordered materials when dynamical effects are dominant, using the echoes as a probe.

 \begin{figure}
\centering
 \includegraphics[scale = 0.15]{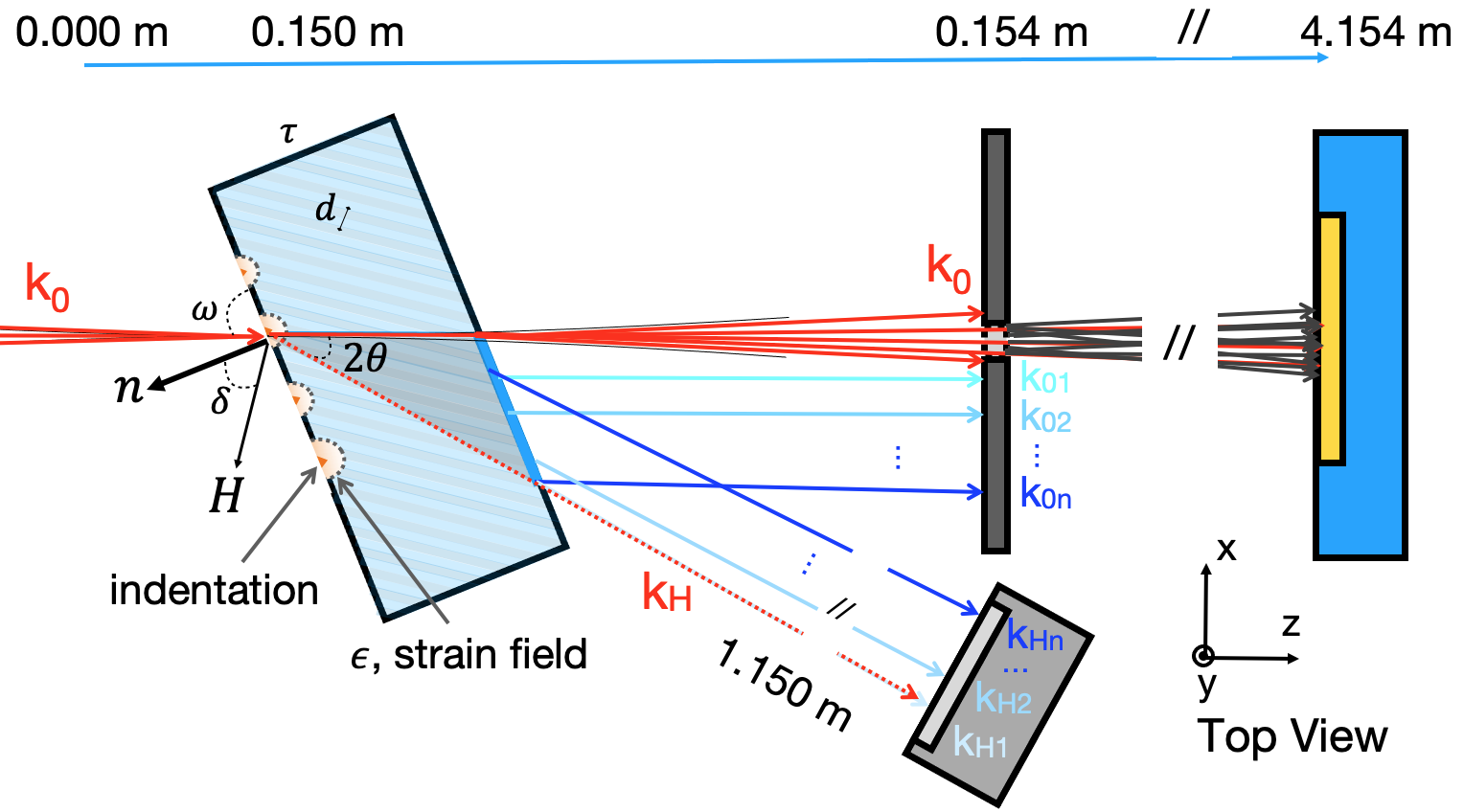}
 \caption{
 Schematics of the diffraction setup. 
 The focused X-ray beam with wavelength $\lambda$ and wave vector $\mathbf{k_0}$, $|\mathbf{k_0}| = 2 \pi/\lambda$, impinges on the Si sample oriented with a family of atomic planes (with spacing $d$) in diffraction conditions, $\mathbf{k_0}+\mathbf{H} = \mathbf{k_H}$, with $\lvert\mathbf{H}\rvert=2\pi /d$.
 The photons are diffracted at an angle $2\theta$ along $\mathbf{k_H}$.
 Echoes in the forward (diffraction) direction are indicated with their wave-vectors $\mathbf{k_{0i}}$ ($\mathbf{k_{Hi}}$).
 The position of the indents on the sample, the pinhole used to scan the wave-front and the two pixel detectors, one in forward and one in diffraction direction, are shown.
}\label{fig:geometry}
\end{figure}

The samples used in this study are two Si$(100)$ wafers of $\SI{100}{\micro\meter}$ thickness.
On the surface of one of them, a series of nano-indents are performed using an Alemnis nanoindenter inside a Zeiss Leo Ultra 55 FEG scanning electron miroscope -SEM, as shown in Fig.~S1 of the SM \cite{SM20}.
Each indent produces a strain field that propagates radially from the indent with an amplitude that decreases exponentially with distance \cite{Reuber14, Liu14}. 
The Si samples are measured with X-rays using a Laue geometry, with the Si(111) planes in diffraction conditions, so that the forward diffraction (FD) signal can be measured, as described in our previous work \cite{ARF18,ARF20}.
A  schematics  of  the  sample  geometry  is shown in Fig.~\ref{fig:geometry}.
The FD signal can be only described using dynamical diffraction theory \cite{Batter64,Zacha1945,Shvy12}, 
which includes all X-ray interactions within the crystal unit, beyond the diffraction of the incident beam with the directly illuminated volume.
Due to the crystal long-range order, the diffracted photons see more lattice planes also in diffraction condition.
Therefore, the photons can be diffracted multiple times
both in the diffracted and forward. 
The diffraction inside the crystal occurs in the volume enclosed by the beam size in the y direction and the gray triangle in Fig.~\ref{fig:geometry} in the x-z plane.
At the crystal exit surface all diffracted waves interfere generating the transverse displaced echoes \cite{Shvy12,ARF18}, both in the forward and diffracted directions, with wave vectors $\mathbf{k_{0i}}$ and $\mathbf{k_{Hi}}$, respectively, with $i = 1 ,2,...$ the number of the X-ray beams generated.
Due to the different path the waves follow inside the crystal, they accumulate a temporal delay with respect to each other \cite{ARF18,Shvy12}.
The time delay $\Delta t$ and the transverse displacement $\Delta x$ associated to each beam are related linearly \cite{Shvy12} by
 \begin{equation}
 \label{eq:deltax_deltat}
 \Delta x= c\cot(\theta)\Delta t
 \end{equation}
where $c$ is the speed of light and $\theta$ is the Bragg angle
\footnote{From Bragg's law $2d \sin \theta=n\lambda$}.

The measurements are performed at the beamline NanoMAX using a photon energy of $\SI{8}{\kilo\electronvolt}$ ($\lambda=\SI{0.155}{\nano\meter}$) selected by a Si$(111)$  monochromator.
At this energy the flux at the sample is of $10^{11}$~ph/s and the Kirkpatrick-Baez (KB) mirrors produce a highly coherent X-ray focused beam with $\sim\SI{110}{\nano\meter}$ waist  and \SI{250}{\micro\meter} focal depth, with a divergence of $\sim\SI{1.2}{\milli\radian}$ in both horizontal and vertical directions \cite{Bjorling2020}. 
The Si samples are mounted on a scanning stage in the KB focal plane. 
A $\SI{3}{\micro\meter}$ diameter pinhole is mounted on a second piezo scanning stage $\SI{3.68}{\milli\meter}$ downstream the sample along the beam propagation axis, as shown schematically in Fig.~\ref{fig:geometry}.
The asymmetric Si$(111)$ reflection, $\theta_{111}=\SI{14.3}{\degree}$, is chosen to match the $\sim\SI{1}{\electronvolt}$ monochromator bandwidth.
For Laue symmetric geometry the extinction length of Si$(111)$ at $\SI{8}{\kilo\electronvolt}$ is $\SI{18.54}{\micro\meter}$ and the absorption depth is $\SI{220.38}{\micro\meter}$.
The detector, a photon counting Merlin system with a 512$\times$512 array of pixels with $\SI{55}{\micro\meter}$ edge size, is placed $\SI{4.5}{\meter}$ downstream the sample, to satisfy the ptychographic sampling requirement from the pinhole at the energy used.
A He filled flight tube is used to minimise air scattering between pinhole and Merlin detector. 
A Pilatus 100K detector is placed in the horizontal scattering plane to optimise the Si(111) reflection.  

\begin{figure}
\centering
 \includegraphics[scale = 0.23]{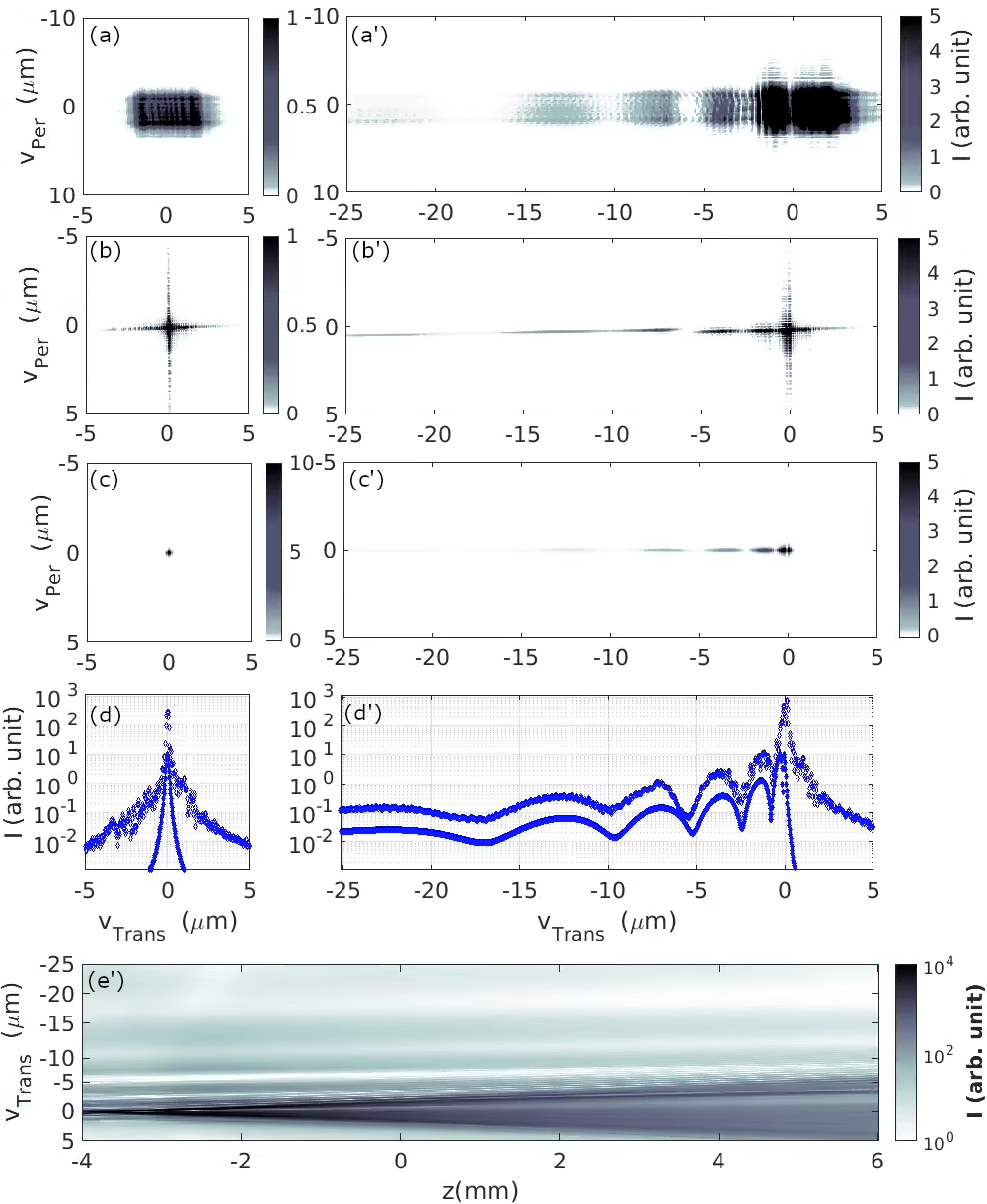}
 \caption{
 Intensity of the FD wave-front from strain-free Si wafer at two angles: (left) $1$ degree away -echoes are absent,  and (right) at the diffraction angle.
 (a,a') Reconstruction at the pinhole plane. 
 (b,b') Reconstruction propagated to the KB focus. 
 (c,c') Simulation at the focus. 
 (d,d') Line-cuts through (a,a') and (b,b'). The simulated signal, lower line, is offset for clarity.
 (e') Propagation of (a') along the beam direction z up to the sample position, illustrating the divergence of the transmitted beam and the parallel echoes. 
 }
 \label{fig:Reconst_prop_sim}
\end{figure}

The intensity of the forward wave-front is measured with tele-ptychography scans \cite{Tsai26}, for both samples and both in and out of the diffraction condition.
Tele-ptychography measurements are performed by scanning the pinhole in the x-y plane, over an area of 18$\times$$\SI{36}{\micro\meter}$$^2$ for the strain free sample and 18$\times$$\SI{27}{\micro\meter}$$^2$ for the nano-indented sample, with a step size of $\SI{0.5}{\micro\meter}$.
In order to counteract possible drifts of the setup, the data-set is acquired over eight separated scans, each covering an area of 10$\times$$\SI{10}{\micro\meter}$$^2$, with $\SI{1}{\micro\meter}$ overlap between adjacent areas, as shown in Fig.~S3 of SM \cite{SM20}. 
Data reconstruction is performed with a tailored routine \cite{Wakonig20} that combines 500 iterations of the difference map algorithm \cite{Thibault09} without updating the reconstruction of the pinhole for the first 499 iterations and 1500 iterations of a maximum likelihood refinement \cite{Thibault12} with update of both wavefront and pinhole.
This routine applies a common refinement of the wavefront in the overlapping areas, while the refinement of the pinhole is independent for the 8 scans \cite{Guizar14}.
In the reconstruction we use the pinhole obtained with a test specimen as initial guess, see Fig.~S4 of the SM \cite{SM20}.
 
The tele-ptychography reconstructions return the amplitude and phase of the forward diffracted wave field and the transmitted beam at the pinhole position (cf.~Fig.~S5 of the SM \cite{SM20}).
Figure \ref{fig:Reconst_prop_sim} (a,a') show the forward diffracted intensity (the square of the retrieved amplitude) from the strain-free sample out and in diffraction conditions, respectively. 
Echoes are indeed only seen in diffraction. 
These measurements reproduce previous results obtained with direct measurements \cite{ARF18} validating  our approach. 
The divergence of the focused beam is not propagated through the echoes in the diffraction plane, but is only visible in the direction perpendicular to it.
In the horizontal plane, indeed, the crystal acts as a X-ray beam filter diffracting only the photons with a divergence comparable with the Darwin width of the selected diffraction peak, which in our case is of $\SI{100}{\nano\radian}$  \cite{Zacha1945}.
In the vertical plane, this filter effect does not apply (cf.~Fig.~\ref{fig:Reconst_prop_sim}(a')).
Nevertheless, the transmitted beam preserves the full divergence of the incoming beam, overlapping with the echoes at large distance from the sample.
Figure \ref{fig:Reconst_prop_sim}(b,b') shows the diffracted intensities of (a,a') at the sample plane, obtained by the propagation of the retrieved wave-field \cite{Tsai26,verezhak2020}.
The sample plane coincides with the location of the X-ray beam focus, and therefore provides the best possible resolution of the forward diffracted field.
Here, we note that there is no strict correspondence, in real space, between the position of the sample in the focus of the beam, its thickness, unintentionally comparable to the beam focal depth, and the propagation of the echoes in the focal plane.
Indeed, the diffracted field measured is produced at the sample exit surface, and the back-propagation of the wavefield beyond this point does not reflect any physical reality. 
The pixel size obtained for the reconstruction is $\SI{32}{\nano\meter}$, while the resolution is estimated to be $\SI{55}{\nano\meter}$, using Fourier ring correlation \cite{Heel05} for two different scans of the forward beam out of diffraction conditions (see Fig.~S6 in SM \cite{SM20}).
Figures~\ref{fig:Reconst_prop_sim}(c,c') show the simulation of the intensity of the forward diffracted field at the sample position for the configuration used, i.e.~a $\SI{100}{\micro\meter}$ thick Si crystal at the asymmetric $(111)$ reflection, using a horizontal diffraction geometry and a strain-free model, as described in \cite{ARF18}.
 The agreement with simulations is more clearly seen in the line-cuts shown in Fig.~\ref{fig:Reconst_prop_sim}(d,d'), where the data are offset for clarity. 
 Figure \ref{fig:Reconst_prop_sim}(e') shows a map of the intensity of the echoes along the beam propagation direction. 
 Due to the beam divergence, already at a distance of $\SI{2}{\milli\meter}$ from the sample surface, the echoes are overlapping with the divergent transmitted beam.
 Moreover, the echoes are parallel to the beam propagation direction, as expected.

\begin{figure}
 \centering
 \includegraphics[scale = 0.3]{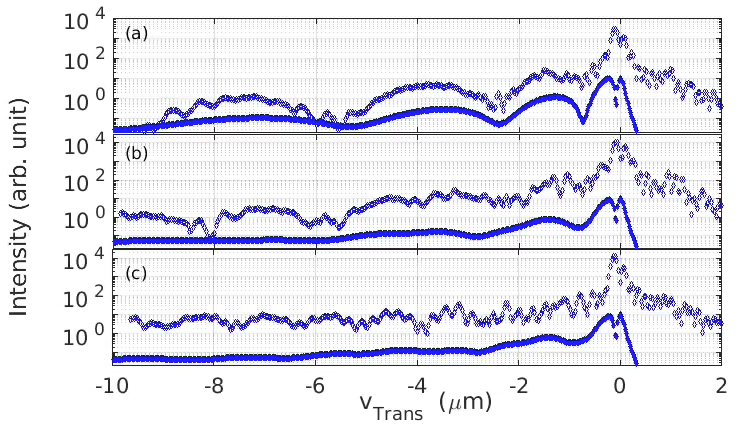}
 \caption{
 Line-cuts through the retrieved (empty dots) and simulated (filled dots) echoes from the indented Si sample in regions with different strain state, cf. text for details.
 Intensities are offset for clarity. 
  }
 \label{fig:Strain-profile}
\end{figure}

From the indented sample data are collected at different positions to compare the FD wave-fronts produced by different strain fields. The dynamical diffraction goes far beyond the volume affected by the indent-induced strain.
As a consequence the echoes are still produced with a large contribution from the strain-free part of the Si crystal.
Simulations of the expected echoes distribution are performed with our code \cite{ARF18,ARF20} that we have adapted to model strained crystals following the work in Ref.~\cite{Lings06,Takagi62,Kato59}.
The model assumes that the indent produces an isotropic strain field of amplitude proportional to the load, which decreases with distance from the indent following an inverse exponential law.
Numerically this is achieved by slicing the crystal in $\SI{20}{\nano\meter}$ layers parallel to the surface, each with a different value of lattice parameter following the exponential decay. This numerical step, necessary to reduce the complexity of the calculations, approximates the radial to a linear decay.
Despite this approximation introduces a variation of a factor between 1 and 2 of the effective decay length X-rays experience along all possible paths defined by the dynamical diffraction process, it still provides good correspondence between data and simulations.
We expect that improving the model to reflect a true radial decay will increase this correspondence.
Finally, our model also disregards the strain anisotropy expected along different crystallographic directions, due to different elastic constants. This should be included in a future model.   
 
\begin{figure}
\centering
 \includegraphics[scale = 0.30]{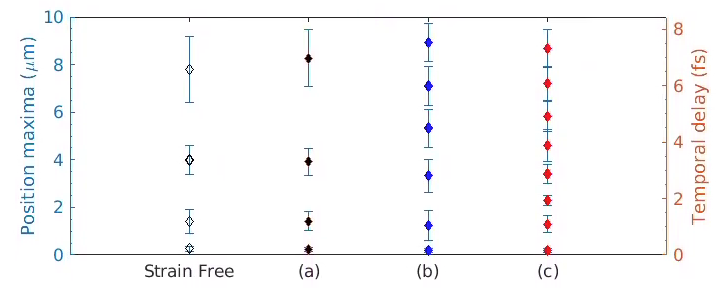}
 \caption{
(Left) Transverse displacement of the echoes along the first $\SI{8}{\micro\meter}$ of the reconstruction with respect to the transmitted beam as extracted from Fig.~S11 in SM at \cite{SM20} and (right) calculated delays from equation (\ref{eq:deltax_deltat}). 
For the strain-free sample and the indented sample at the positions (a), (b) and (c) as presented in Fig. \ref{fig:Strain-profile}.
The bars correspond to the half width half maximum of the respective echoes
}\label{fig:table_time}
\end{figure}

 Figure \ref{fig:Strain-profile}(a), (b) and (c) show the results from the inversion of the data collected  respectively at \SI{1}{\milli\m} from the indented area, and close to two nano-indents with loads $\SI{25}{\milli\newton}$ and $\SI{75}{\milli\newton}$. 
 The FD signal retrieved shows an increase of the number of echoes and a decrease of their mutual distance for increasing indentation strength. 
 At higher strain, the echoes tend to merge into a continuum.
 Simulations of the dynamical diffraction signal using models with different strain fields provide a rather good  agreement with data.  Tele-ptychography reconstructions from this sample and the respective simulations are shown in Fig.~S7-S10 of the SM.
 More accurate fitting routines will be implemented in the future.
 However, the position of the echoes are easily extracted via algorithms to find local maxima, as shown in Fig.~S11.
 New echoes with shorter time delays arise in the strained areas, confirming this experimental approach to observe the strain-tailored FD maxima involved in this ultra-fast process.
 Their time delay, visualised in Fig.~S12 of SM, can be calculated using  Eq.(\ref{eq:deltax_deltat}) and is shown in Fig.~\ref{fig:table_time}.
 The strain amplitude (expressed as fractional variation of the lattice parameter) and the decay length of the exponential that best reproduces the data for the three regions (see Fig.~S9 in SM at \cite{SM20}) are: 
 \begin{itemize}[itemsep=-0.3em]
  \item $5\cdot10^{-5}$ over $\SI{2}{\micro\meter}$, profile (a)
  \item $10^{-4}$ over $\SI{7}{\micro\meter}$, profile (b)
  \item $10^{-4}$ over $\SI{10}{\micro\meter}$, profile (c)
\end{itemize}
These findings are in agreement with results from literature \cite{Li2015}.

 These pioneering results demonstrate the extreme sensitivity of dynamical diffraction effects to localised strain fields of small amplitude. 
 Echoes are produced from a volume extending throughout the full sample thickness, of which the strained area represents a mere $10\%$. 
 Further improvement of the model will include the implementation of a radial decay of the strain field and eventually the crystal elastic anisotropy.
 A natural progression of this research is measuring echoes in the diffraction direction, where the spatio-temporal signal does not suffer from the absorption effects, as in FD.
 As presented in Fig.~S13 in SM at \cite{SM20}, the echoes probe the crystal in a more homogeneous manner, which allows to retrieve more clearly the information from lattice deformations.
 
 We have presented an experimental study of the dynamical diffraction produced by a Si wafer in transmission geometry, in presence of a localised strain. We have shown that strain can be used to finely tune time delay of the echoes, in a way we are able to model. We achieve a high resolution in both, sample and detection space with the use of a focused beam combined with a tele-ptychography approach. 
 With the use of a nanobeam, we effectively increase the sensitivity of our measurement to strain gradients and sample heterogeneity by reducing the illuminated volume. 
 Our approach provides a sensitive and efficient way to analyse the impact of strain of different amplitudes onto dynamical diffraction effects from perfect crystals.
 Conversely, understanding how dynamical diffraction is affected by the presence of strain in otherwise perfectly ordered crystals offers a new and visibly more sensitive tool for the characterisation of strain and defects in crystalline materials.
 This work provides insights on the dynamical effects arising in coherent imaging experiments involving  high atomic-number crystals, such as Ni, Au, InSb or CdTe, for which the extinction length reduces to the \SI{}{\micro\meter} and sub-\SI{}{\micro\meter} level, and contributes to the landscape of emerging experimental and analysis approaches developed to model them.
 It is important to mention that the presence from the echoes in the diffracted signal could be distorting the results in temporal ultrafast studies.
 
Finally, being able to predict and control the behaviour of dynamical diffraction in presence of strain opens the fascinating possibility of strain-tailoring ultrafast optics.
This might include the generation of multi bunch sources from single ultrafast pulses or the production of tailored crystals for split-and-delay lines with full control of the timing and width of the generated pulses with respect to the incoming pulse.
 
 We acknowledge MAX IV Laboratory for beamtime under Proposal 20180253. MAX IV is supported by the Swedish Research council, Vinnova and Formas under contract 2018-07152, 2018-04969 and 2019-02496, respectively.
 NanoMAX staff is acknowledged for support during the preparation and the execution of the experiment.
 M.V. acknowledges funding by European Union's Horizon 2020 research and innovation program under the Marie Sk\l{}odowska-Curie grant agreement No 701647, and the SNSF grant No 200021L\_169753.
 We thank Zdenek Matej, Manuel Guizar-Sicairos, Bill Pedrini, Virginie Chamard and Kenneth Finkelstein for useful discussions.

 A.R.F and DC conceptualized the work; H.S.I. and M.H.C. prepared the samples; A.R.F., D.C, M.H.C., M.V. and A.D. planned the experiment;  A.R.F., D.C and A.D. performed the experiment; A.R.F. analysed the data with contributions of A.D., K.W. and M.V.; A.R.F. performed the simulations; D.C. and A.R.F. wrote the manuscript with the contribution of all authors.

 \bibliography{TeleptychoNanoMAX}

\begin{thebibliography}{46}%
\makeatletter
\providecommand \@ifxundefined [1]{%
 \@ifx{#1\undefined}
}%
\providecommand \@ifnum [1]{%
 \ifnum #1\expandafter \@firstoftwo
 \else \expandafter \@secondoftwo
 \fi
}%
\providecommand \@ifx [1]{%
 \ifx #1\expandafter \@firstoftwo
 \else \expandafter \@secondoftwo
 \fi
}%
\providecommand \natexlab [1]{#1}%
\providecommand \enquote  [1]{``#1''}%
\providecommand \bibnamefont  [1]{#1}%
\providecommand \bibfnamefont [1]{#1}%
\providecommand \citenamefont [1]{#1}%
\providecommand \href@noop [0]{\@secondoftwo}%
\providecommand \href [0]{\begingroup \@sanitize@url \@href}%
\providecommand \@href[1]{\@@startlink{#1}\@@href}%
\providecommand \@@href[1]{\endgroup#1\@@endlink}%
\providecommand \@sanitize@url [0]{\catcode `\\12\catcode `\$12\catcode
  `\&12\catcode `\#12\catcode `\^12\catcode `\_12\catcode `\%12\relax}%
\providecommand \@@startlink[1]{}%
\providecommand \@@endlink[0]{}%
\providecommand \url  [0]{\begingroup\@sanitize@url \@url }%
\providecommand \@url [1]{\endgroup\@href {#1}{\urlprefix }}%
\providecommand \urlprefix  [0]{URL }%
\providecommand \Eprint [0]{\href }%
\providecommand \doibase [0]{https://doi.org/}%
\providecommand \selectlanguage [0]{\@gobble}%
\providecommand \bibinfo  [0]{\@secondoftwo}%
\providecommand \bibfield  [0]{\@secondoftwo}%
\providecommand \translation [1]{[#1]}%
\providecommand \BibitemOpen [0]{}%
\providecommand \bibitemStop [0]{}%
\providecommand \bibitemNoStop [0]{.\EOS\space}%
\providecommand \EOS [0]{\spacefactor3000\relax}%
\providecommand \BibitemShut  [1]{\csname bibitem#1\endcsname}%
\let\auto@bib@innerbib\@empty
\bibitem [{\citenamefont {Tavares~et al.}(2018)}]{MAXIV18}%
  \BibitemOpen
  \bibfield  {author} {\bibinfo {author} {\bibfnamefont {P.~F.}\ \bibnamefont
  {Tavares~et al.}},\ }\bibfield  {title} {\bibinfo {title} {Commissioning and
  first-year operational results of the \ce{MAX IV 3 GeV} ring},\ }\href@noop
  {} {\bibfield  {journal} {\bibinfo  {journal} {J. Synchrotron Rad.}\ }\textbf
  {\bibinfo {volume} {25}},\ \bibinfo {pages} {1291} (\bibinfo {year}
  {2018})}\BibitemShut {NoStop}%
\bibitem [{\citenamefont {Decking~et al.}(2020)}]{EUXFEL20}%
  \BibitemOpen
  \bibfield  {author} {\bibinfo {author} {\bibfnamefont {W.}~\bibnamefont
  {Decking~et al.}},\ }\bibfield  {title} {\bibinfo {title} {A
  \ce{MHz}-repetition-rate hard x-ray free-electron laser driven by a
  superconducting linear accelerator},\ }\href@noop {} {\bibfield  {journal}
  {\bibinfo  {journal} {Nat. Photonics}\ }\textbf {\bibinfo {volume} {14}},\
  \bibinfo {pages} {391} (\bibinfo {year} {2020})}\BibitemShut {NoStop}%
\bibitem [{\citenamefont {Glownia~et al.}(2010)}]{LCLS10}%
  \BibitemOpen
  \bibfield  {author} {\bibinfo {author} {\bibfnamefont {J.~M.}\ \bibnamefont
  {Glownia~et al.}},\ }\bibfield  {title} {\bibinfo {title} {Time-resolved
  pump-probe experiments at the \ce{LCLS}},\ }\href@noop {} {\bibfield
  {journal} {\bibinfo  {journal} {Opt. Express}\ }\textbf {\bibinfo {volume}
  {18}},\ \bibinfo {pages} {17620} (\bibinfo {year} {2010})}\BibitemShut
  {NoStop}%
\bibitem [{\citenamefont {Pietsch}\ \emph {et~al.}(2004)\citenamefont
  {Pietsch}, \citenamefont {Holy},\ and\ \citenamefont {Baumbach}}]{Vaclav04}%
  \BibitemOpen
  \bibfield  {author} {\bibinfo {author} {\bibfnamefont {U.}~\bibnamefont
  {Pietsch}}, \bibinfo {author} {\bibfnamefont {V.}~\bibnamefont {Holy}},\ and\
  \bibinfo {author} {\bibfnamefont {T.}~\bibnamefont {Baumbach}},\ }\href@noop
  {} {\emph {\bibinfo {title} {High-Resolution X-Ray Scattering: From Thin
  Films to Lateral Nanostructures}}}\ (\bibinfo  {publisher} {Springer-Verlad
  New York , LLC},\ \bibinfo {year} {2004})\BibitemShut {NoStop}%
\bibitem [{\citenamefont {Thibault}\ \emph {et~al.}(2008)\citenamefont
  {Thibault}, \citenamefont {Dierolf}, \citenamefont {Menzel}, \citenamefont
  {Bunk}, \citenamefont {David},\ and\ \citenamefont {Pfeiffer}}]{Thibault379}%
  \BibitemOpen
  \bibfield  {author} {\bibinfo {author} {\bibfnamefont {P.}~\bibnamefont
  {Thibault}}, \bibinfo {author} {\bibfnamefont {M.}~\bibnamefont {Dierolf}},
  \bibinfo {author} {\bibfnamefont {A.}~\bibnamefont {Menzel}}, \bibinfo
  {author} {\bibfnamefont {O.}~\bibnamefont {Bunk}}, \bibinfo {author}
  {\bibfnamefont {C.}~\bibnamefont {David}},\ and\ \bibinfo {author}
  {\bibfnamefont {F.}~\bibnamefont {Pfeiffer}},\ }\bibfield  {title} {\bibinfo
  {title} {High-resolution scanning x-ray diffraction microscopy},\ }\href
  {https://doi.org/10.1126/science.1158573} {\bibfield  {journal} {\bibinfo
  {journal} {Science}\ }\textbf {\bibinfo {volume} {321}},\ \bibinfo {pages}
  {379} (\bibinfo {year} {2008})},\ \Eprint
  {https://arxiv.org/abs/https://science.sciencemag.org/content/321/5887/379.full.pdf}
  {https://science.sciencemag.org/content/321/5887/379.full.pdf} \BibitemShut
  {NoStop}%
\bibitem [{\citenamefont {Ulvestad~et al.}(2014)}]{Ulvestad2014}%
  \BibitemOpen
  \bibfield  {author} {\bibinfo {author} {\bibfnamefont {A.}~\bibnamefont
  {Ulvestad~et al.}},\ }\bibfield  {title} {\bibinfo {title} {{Single Particle
  Nanomechanics in Operando Batteries via Lensless Strain Mapping}},\
  }\href@noop {} {\bibfield  {journal} {\bibinfo  {journal} {Nano Lett.}\
  }\textbf {\bibinfo {volume} {14}},\ \bibinfo {pages} {5123} (\bibinfo {year}
  {2014})}\BibitemShut {NoStop}%
\bibitem [{\citenamefont {Singer~et al.}(2018)}]{Singer2018}%
  \BibitemOpen
  \bibfield  {author} {\bibinfo {author} {\bibfnamefont {A.}~\bibnamefont
  {Singer~et al.}},\ }\bibfield  {title} {\bibinfo {title} {Nucleation of
  dislocations and their dynamics in layered oxide cathode materials during
  battery charging},\ }\href@noop {} {\bibfield  {journal} {\bibinfo  {journal}
  {Nature Energy}\ }\textbf {\bibinfo {volume} {3}},\ \bibinfo {pages} {641}
  (\bibinfo {year} {2018})}\BibitemShut {NoStop}%
\bibitem [{\citenamefont {Godard~et al.}(2011)}]{Godard11}%
  \BibitemOpen
  \bibfield  {author} {\bibinfo {author} {\bibfnamefont {P.}~\bibnamefont
  {Godard~et al.}},\ }\bibfield  {title} {\bibinfo {title} {Three-dimensional
  high-resolution quantitative microscopy of extended crystals},\ }\href@noop
  {} {\bibfield  {journal} {\bibinfo  {journal} {Nature Communications}\
  }\textbf {\bibinfo {volume} {2}},\ \bibinfo {pages} {568} (\bibinfo {year}
  {2011})}\BibitemShut {NoStop}%
\bibitem [{\citenamefont {Hill~et al.}(2018)}]{Hill18}%
  \BibitemOpen
  \bibfield  {author} {\bibinfo {author} {\bibfnamefont {M.~O.}\ \bibnamefont
  {Hill~et al.}},\ }\bibfield  {title} {\bibinfo {title} {Measuring
  three-dimensional strain and structural defects in a single \ce{InGaAs}
  nanowire using coherent x-ray multiangle bragg projection ptychography},\
  }\href@noop {} {\bibfield  {journal} {\bibinfo  {journal} {Nano Letters}\
  }\textbf {\bibinfo {volume} {18}},\ \bibinfo {pages} {811} (\bibinfo {year}
  {2018})}\BibitemShut {NoStop}%
\bibitem [{\citenamefont {Hruszkewycz~et al.}(2017)}]{Hruszkewycz2017}%
  \BibitemOpen
  \bibfield  {author} {\bibinfo {author} {\bibfnamefont {S.~O.}\ \bibnamefont
  {Hruszkewycz~et al.}},\ }\bibfield  {title} {\bibinfo {title}
  {High-resolution three-dimensional structural microscopy by single-angle
  bragg ptychography},\ }\href {https://doi.org/10.1038/nmat4798} {\bibfield
  {journal} {\bibinfo  {journal} {Nature Materials}\ }\textbf {\bibinfo
  {volume} {16}},\ \bibinfo {pages} {244} (\bibinfo {year} {2017})}\BibitemShut
  {NoStop}%
\bibitem [{\citenamefont {Chamard}\ \emph {et~al.}(2010)\citenamefont
  {Chamard}, \citenamefont {Stangl}, \citenamefont {Carbone}, \citenamefont
  {Diaz}, \citenamefont {Chen}, \citenamefont {Alfonso}, \citenamefont
  {Mocuta},\ and\ \citenamefont {Metzger}}]{Chamard10}%
  \BibitemOpen
  \bibfield  {author} {\bibinfo {author} {\bibfnamefont {V.}~\bibnamefont
  {Chamard}}, \bibinfo {author} {\bibfnamefont {J.}~\bibnamefont {Stangl}},
  \bibinfo {author} {\bibfnamefont {G.}~\bibnamefont {Carbone}}, \bibinfo
  {author} {\bibfnamefont {A.}~\bibnamefont {Diaz}}, \bibinfo {author}
  {\bibfnamefont {G.}~\bibnamefont {Chen}}, \bibinfo {author} {\bibfnamefont
  {C.}~\bibnamefont {Alfonso}}, \bibinfo {author} {\bibfnamefont
  {C.}~\bibnamefont {Mocuta}},\ and\ \bibinfo {author} {\bibfnamefont {T.~H.}\
  \bibnamefont {Metzger}},\ }\bibfield  {title} {\bibinfo {title}
  {Three-dimensional x-ray fourier transform holography: The bragg case},\
  }\href {https://doi.org/10.1103/PhysRevLett.104.165501} {\bibfield  {journal}
  {\bibinfo  {journal} {Phys. Rev. Lett.}\ }\textbf {\bibinfo {volume} {104}},\
  \bibinfo {pages} {165501} (\bibinfo {year} {2010})}\BibitemShut {NoStop}%
\bibitem [{\citenamefont {Zhang}\ \emph {et~al.}(2018)\citenamefont {Zhang},
  \citenamefont {Dufresne},\ and\ \citenamefont {Sandy}}]{Zhang18}%
  \BibitemOpen
  \bibfield  {author} {\bibinfo {author} {\bibfnamefont {Q.}~\bibnamefont
  {Zhang}}, \bibinfo {author} {\bibfnamefont {E.~M.}\ \bibnamefont
  {Dufresne}},\ and\ \bibinfo {author} {\bibfnamefont {A.~R.}\ \bibnamefont
  {Sandy}},\ }\bibfield  {title} {\bibinfo {title} {Dynamics in hard condensed
  matter probed by x-ray photon correlation spectroscopy: Present and beyond},\
  }\href {https://doi.org/https://doi.org/10.1016/j.cossms.2018.06.002}
  {\bibfield  {journal} {\bibinfo  {journal} {Current Opinion in Solid State
  and Materials Science}\ }\textbf {\bibinfo {volume} {22}},\ \bibinfo {pages}
  {202} (\bibinfo {year} {2018})},\ \bibinfo {note} {advanced characterization
  of nanomaterials}\BibitemShut {NoStop}%
\bibitem [{\citenamefont {Jacques}\ \emph {et~al.}(2016)\citenamefont
  {Jacques}, \citenamefont {Laulhé}, \citenamefont {Moisan}, \citenamefont
  {Ravy},\ and\ \citenamefont {Le~Bolloc’h}}]{Jacques16}%
  \BibitemOpen
  \bibfield  {author} {\bibinfo {author} {\bibfnamefont {V.}~\bibnamefont
  {Jacques}}, \bibinfo {author} {\bibfnamefont {C.}~\bibnamefont {Laulhé}},
  \bibinfo {author} {\bibfnamefont {N.}~\bibnamefont {Moisan}}, \bibinfo
  {author} {\bibfnamefont {S.}~\bibnamefont {Ravy}},\ and\ \bibinfo {author}
  {\bibfnamefont {D.}~\bibnamefont {Le~Bolloc’h}},\ }\bibfield  {title}
  {\bibinfo {title} {Laser-induced charge-density-wave transient depinning in
  chromium},\ }\href {https://doi.org/10.1103/PhysRevLett.117.156401}
  {\bibfield  {journal} {\bibinfo  {journal} {Phys. Rev. Lett.}\ }\textbf
  {\bibinfo {volume} {117}},\ \bibinfo {pages} {156401} (\bibinfo {year}
  {2016})}\BibitemShut {NoStop}%
\bibitem [{\citenamefont {Schropp}\ \emph {et~al.}(2010)\citenamefont
  {Schropp}, \citenamefont {Boye}, \citenamefont {Feldkamp}, \citenamefont
  {Hoppe}, \citenamefont {Patommel}, \citenamefont {Samberg}, \citenamefont
  {Stephan}, \citenamefont {Giewekemeyer}, \citenamefont {Wilke}, \citenamefont
  {Salditt}, \citenamefont {Gulden}, \citenamefont {Mancuso}, \citenamefont
  {Vartanyants}, \citenamefont {Weckert}, \citenamefont {Schöder},
  \citenamefont {Burghammer},\ and\ \citenamefont {Schroer}}]{Schropp2010}%
  \BibitemOpen
  \bibfield  {author} {\bibinfo {author} {\bibfnamefont {A.}~\bibnamefont
  {Schropp}}, \bibinfo {author} {\bibfnamefont {P.}~\bibnamefont {Boye}},
  \bibinfo {author} {\bibfnamefont {J.~M.}\ \bibnamefont {Feldkamp}}, \bibinfo
  {author} {\bibfnamefont {R.}~\bibnamefont {Hoppe}}, \bibinfo {author}
  {\bibfnamefont {J.}~\bibnamefont {Patommel}}, \bibinfo {author}
  {\bibfnamefont {D.}~\bibnamefont {Samberg}}, \bibinfo {author} {\bibfnamefont
  {S.}~\bibnamefont {Stephan}}, \bibinfo {author} {\bibfnamefont
  {K.}~\bibnamefont {Giewekemeyer}}, \bibinfo {author} {\bibfnamefont {R.~N.}\
  \bibnamefont {Wilke}}, \bibinfo {author} {\bibfnamefont {T.}~\bibnamefont
  {Salditt}}, \bibinfo {author} {\bibfnamefont {J.}~\bibnamefont {Gulden}},
  \bibinfo {author} {\bibfnamefont {A.~P.}\ \bibnamefont {Mancuso}}, \bibinfo
  {author} {\bibfnamefont {I.~A.}\ \bibnamefont {Vartanyants}}, \bibinfo
  {author} {\bibfnamefont {E.}~\bibnamefont {Weckert}}, \bibinfo {author}
  {\bibfnamefont {S.}~\bibnamefont {Schöder}}, \bibinfo {author}
  {\bibfnamefont {M.}~\bibnamefont {Burghammer}},\ and\ \bibinfo {author}
  {\bibfnamefont {C.~G.}\ \bibnamefont {Schroer}},\ }\bibfield  {title}
  {\bibinfo {title} {Hard x-ray nanobeam characterization by coherent
  diffraction microscopy},\ }\href {https://doi.org/10.1063/1.3332591}
  {\bibfield  {journal} {\bibinfo  {journal} {Applied Physics Letters}\
  }\textbf {\bibinfo {volume} {96}},\ \bibinfo {pages} {091102} (\bibinfo
  {year} {2010})},\ \Eprint
  {https://arxiv.org/abs/https://doi.org/10.1063/1.3332591}
  {https://doi.org/10.1063/1.3332591} \BibitemShut {NoStop}%
\bibitem [{\citenamefont {Vila-Comamala}\ \emph {et~al.}(2011)\citenamefont
  {Vila-Comamala}, \citenamefont {Diaz}, \citenamefont {Guizar-Sicairos},
  \citenamefont {Mantion}, \citenamefont {Kewish}, \citenamefont {Menzel},
  \citenamefont {Bunk},\ and\ \citenamefont {David}}]{Vila2011}%
  \BibitemOpen
  \bibfield  {author} {\bibinfo {author} {\bibfnamefont {J.}~\bibnamefont
  {Vila-Comamala}}, \bibinfo {author} {\bibfnamefont {A.}~\bibnamefont {Diaz}},
  \bibinfo {author} {\bibfnamefont {M.}~\bibnamefont {Guizar-Sicairos}},
  \bibinfo {author} {\bibfnamefont {A.}~\bibnamefont {Mantion}}, \bibinfo
  {author} {\bibfnamefont {C.~M.}\ \bibnamefont {Kewish}}, \bibinfo {author}
  {\bibfnamefont {A.}~\bibnamefont {Menzel}}, \bibinfo {author} {\bibfnamefont
  {O.}~\bibnamefont {Bunk}},\ and\ \bibinfo {author} {\bibfnamefont
  {C.}~\bibnamefont {David}},\ }\bibfield  {title} {\bibinfo {title}
  {Characterization of high-resolution diffractive x-ray optics by
  ptychographic coherent diffractive imaging},\ }\href
  {https://doi.org/10.1364/OE.19.021333} {\bibfield  {journal} {\bibinfo
  {journal} {Opt. Express}\ }\textbf {\bibinfo {volume} {19}},\ \bibinfo
  {pages} {21333} (\bibinfo {year} {2011})}\BibitemShut {NoStop}%
\bibitem [{\citenamefont {Schropp}\ \emph {et~al.}(2013)\citenamefont
  {Schropp}, \citenamefont {Hoppe}, \citenamefont {Meier}, \citenamefont
  {Patommel}, \citenamefont {Seiboth}, \citenamefont {Lee}, \citenamefont
  {Nagler}, \citenamefont {Galtier}, \citenamefont {Arnold}, \citenamefont
  {Zastrau}, \citenamefont {Hastings}, \citenamefont {Nilsson}, \citenamefont
  {Uhlén}, \citenamefont {Vogt}, \citenamefont {Hertz},\ and\ \citenamefont
  {Schroer}}]{Schropp2013}%
  \BibitemOpen
  \bibfield  {author} {\bibinfo {author} {\bibfnamefont {A.}~\bibnamefont
  {Schropp}}, \bibinfo {author} {\bibfnamefont {R.}~\bibnamefont {Hoppe}},
  \bibinfo {author} {\bibfnamefont {V.}~\bibnamefont {Meier}}, \bibinfo
  {author} {\bibfnamefont {J.}~\bibnamefont {Patommel}}, \bibinfo {author}
  {\bibfnamefont {F.}~\bibnamefont {Seiboth}}, \bibinfo {author} {\bibfnamefont
  {H.~J.}\ \bibnamefont {Lee}}, \bibinfo {author} {\bibfnamefont
  {B.}~\bibnamefont {Nagler}}, \bibinfo {author} {\bibfnamefont {E.~C.}\
  \bibnamefont {Galtier}}, \bibinfo {author} {\bibfnamefont {B.}~\bibnamefont
  {Arnold}}, \bibinfo {author} {\bibfnamefont {U.}~\bibnamefont {Zastrau}},
  \bibinfo {author} {\bibfnamefont {J.~B.}\ \bibnamefont {Hastings}}, \bibinfo
  {author} {\bibfnamefont {D.}~\bibnamefont {Nilsson}}, \bibinfo {author}
  {\bibfnamefont {F.}~\bibnamefont {Uhlén}}, \bibinfo {author} {\bibfnamefont
  {U.}~\bibnamefont {Vogt}}, \bibinfo {author} {\bibfnamefont {H.~M.}\
  \bibnamefont {Hertz}},\ and\ \bibinfo {author} {\bibfnamefont {C.~G.}\
  \bibnamefont {Schroer}},\ }\bibfield  {title} {\bibinfo {title} {Full spatial
  characterization of a nanofocused x-ray free-electron laser beam by
  ptychographic imaging},\ }\href {https://doi.org/10.1038/srep01633}
  {\bibfield  {journal} {\bibinfo  {journal} {Scientific Reports}\ }\textbf
  {\bibinfo {volume} {3}},\ \bibinfo {pages} {1633} (\bibinfo {year}
  {2013})}\BibitemShut {NoStop}%
\bibitem [{\citenamefont {Bj{\"{o}}rling~et al.}(2020)}]{Bjorling2020}%
  \BibitemOpen
  \bibfield  {author} {\bibinfo {author} {\bibfnamefont {A.}~\bibnamefont
  {Bj{\"{o}}rling~et al.}},\ }\bibfield  {title} {\bibinfo {title}
  {{Ptychographic characterization of a coherent nanofocused X-ray beam}},\
  }\href@noop {} {\bibfield  {journal} {\bibinfo  {journal} {Opt. Express}\
  }\textbf {\bibinfo {volume} {28}},\ \bibinfo {pages} {5069} (\bibinfo {year}
  {2020})}\BibitemShut {NoStop}%
\bibitem [{\citenamefont {Zachariasen}(1945)}]{Zacha1945}%
  \BibitemOpen
  \bibfield  {author} {\bibinfo {author} {\bibfnamefont {W.~H.}\ \bibnamefont
  {Zachariasen}},\ }\href@noop {} {\emph {\bibinfo {title} {Theory of X-ray
  Diffraction in Crystals}}}\ (\bibinfo  {publisher} {Dover Publications, INC,
  New York},\ \bibinfo {year} {1945})\BibitemShut {NoStop}%
\bibitem [{\citenamefont {Batterman}\ and\ \citenamefont
  {Cole}(1964)}]{Batter64}%
  \BibitemOpen
  \bibfield  {author} {\bibinfo {author} {\bibfnamefont {B.}~\bibnamefont
  {Batterman}}\ and\ \bibinfo {author} {\bibfnamefont {H.}~\bibnamefont
  {Cole}},\ }\bibfield  {title} {\bibinfo {title} {Dynamical diffraction of
  x-rays by perfect crystals},\ }\href@noop {} {\bibfield  {journal} {\bibinfo
  {journal} {Reviews of Modern Physics}\ }\textbf {\bibinfo {volume} {36}},\
  \bibinfo {pages} {681} (\bibinfo {year} {1964})}\BibitemShut {NoStop}%
\bibitem [{\citenamefont {Authier}(2001)}]{Authi01}%
  \BibitemOpen
  \bibfield  {author} {\bibinfo {author} {\bibfnamefont {A.}~\bibnamefont
  {Authier}},\ }\href@noop {} {\emph {\bibinfo {title} {Dynamical theory of
  X-ray diffraction}}}\ (\bibinfo  {publisher} {Oxford University Press},\
  \bibinfo {year} {2001})\BibitemShut {NoStop}%
\bibitem [{\citenamefont {Amann~et al.}(2012)}]{Amann2012}%
  \BibitemOpen
  \bibfield  {author} {\bibinfo {author} {\bibfnamefont {J.}~\bibnamefont
  {Amann~et al.}},\ }\bibfield  {title} {\bibinfo {title} {Demonstration of
  self-seeding in a hard-x-ray free-electron laser},\ }\href@noop {} {\bibfield
   {journal} {\bibinfo  {journal} {Nature Photonics}\ }\textbf {\bibinfo
  {volume} {6}},\ \bibinfo {pages} {693} (\bibinfo {year} {2012})}\BibitemShut
  {NoStop}%
\bibitem [{\citenamefont {Pateras}\ \emph {et~al.}(2018)\citenamefont
  {Pateras}, \citenamefont {Park}, \citenamefont {Ahn}, \citenamefont {Tilka},
  \citenamefont {Holt}, \citenamefont {Kim}, \citenamefont {Mawst},\ and\
  \citenamefont {Evans}}]{Pateras18}%
  \BibitemOpen
  \bibfield  {author} {\bibinfo {author} {\bibfnamefont {A.}~\bibnamefont
  {Pateras}}, \bibinfo {author} {\bibfnamefont {J.}~\bibnamefont {Park}},
  \bibinfo {author} {\bibfnamefont {Y.}~\bibnamefont {Ahn}}, \bibinfo {author}
  {\bibfnamefont {J.~A.}\ \bibnamefont {Tilka}}, \bibinfo {author}
  {\bibfnamefont {M.~V.}\ \bibnamefont {Holt}}, \bibinfo {author}
  {\bibfnamefont {H.}~\bibnamefont {Kim}}, \bibinfo {author} {\bibfnamefont
  {L.~J.}\ \bibnamefont {Mawst}},\ and\ \bibinfo {author} {\bibfnamefont
  {P.~G.}\ \bibnamefont {Evans}},\ }\bibfield  {title} {\bibinfo {title}
  {Dynamical scattering in coherent hard x-ray nanobeam bragg diffraction},\
  }\href {https://doi.org/10.1103/PhysRevB.97.235414} {\bibfield  {journal}
  {\bibinfo  {journal} {Phys. Rev. B}\ }\textbf {\bibinfo {volume} {97}},\
  \bibinfo {pages} {235414} (\bibinfo {year} {2018})}\BibitemShut {NoStop}%
\bibitem [{\citenamefont {Civita}\ \emph {et~al.}(2018)\citenamefont {Civita},
  \citenamefont {Diaz}, \citenamefont {Bean}, \citenamefont {Shabalin},
  \citenamefont {Gorobtsov}, \citenamefont {Vartanyants},\ and\ \citenamefont
  {Robinson}}]{Civita18}%
  \BibitemOpen
  \bibfield  {author} {\bibinfo {author} {\bibfnamefont {M.}~\bibnamefont
  {Civita}}, \bibinfo {author} {\bibfnamefont {A.}~\bibnamefont {Diaz}},
  \bibinfo {author} {\bibfnamefont {R.~J.}\ \bibnamefont {Bean}}, \bibinfo
  {author} {\bibfnamefont {A.~G.}\ \bibnamefont {Shabalin}}, \bibinfo {author}
  {\bibfnamefont {O.~Y.}\ \bibnamefont {Gorobtsov}}, \bibinfo {author}
  {\bibfnamefont {I.~A.}\ \bibnamefont {Vartanyants}},\ and\ \bibinfo {author}
  {\bibfnamefont {I.~K.}\ \bibnamefont {Robinson}},\ }\bibfield  {title}
  {\bibinfo {title} {Phase modulation due to crystal diffraction by
  ptychographic imaging},\ }\href {https://doi.org/10.1103/PhysRevB.97.104101}
  {\bibfield  {journal} {\bibinfo  {journal} {Phys. Rev. B}\ }\textbf {\bibinfo
  {volume} {97}},\ \bibinfo {pages} {104101} (\bibinfo {year}
  {2018})}\BibitemShut {NoStop}%
\bibitem [{\citenamefont {Shabalin}\ \emph {et~al.}(2017)\citenamefont
  {Shabalin}, \citenamefont {Yefanov}, \citenamefont {Nosik}, \citenamefont
  {Bushuev},\ and\ \citenamefont {Vartanyants}}]{Shabalin17}%
  \BibitemOpen
  \bibfield  {author} {\bibinfo {author} {\bibfnamefont {A.~G.}\ \bibnamefont
  {Shabalin}}, \bibinfo {author} {\bibfnamefont {O.~M.}\ \bibnamefont
  {Yefanov}}, \bibinfo {author} {\bibfnamefont {V.~L.}\ \bibnamefont {Nosik}},
  \bibinfo {author} {\bibfnamefont {V.~A.}\ \bibnamefont {Bushuev}},\ and\
  \bibinfo {author} {\bibfnamefont {I.~A.}\ \bibnamefont {Vartanyants}},\
  }\bibfield  {title} {\bibinfo {title} {Dynamical effects in bragg coherent
  x-ray diffraction imaging of finite crystals},\ }\href
  {https://doi.org/10.1103/PhysRevB.96.064111} {\bibfield  {journal} {\bibinfo
  {journal} {Phys. Rev. B}\ }\textbf {\bibinfo {volume} {96}},\ \bibinfo
  {pages} {064111} (\bibinfo {year} {2017})}\BibitemShut {NoStop}%
\bibitem [{\citenamefont {Gorobtsov}\ and\ \citenamefont
  {Vartanyants}(2016)}]{Gorobtsov16}%
  \BibitemOpen
  \bibfield  {author} {\bibinfo {author} {\bibfnamefont {O.~Y.}\ \bibnamefont
  {Gorobtsov}}\ and\ \bibinfo {author} {\bibfnamefont {I.~A.}\ \bibnamefont
  {Vartanyants}},\ }\bibfield  {title} {\bibinfo {title} {Phase of transmitted
  wave in dynamical theory and quasi-kinematical approximation},\ }\href
  {https://doi.org/10.1103/PhysRevB.93.184107} {\bibfield  {journal} {\bibinfo
  {journal} {Phys. Rev. B}\ }\textbf {\bibinfo {volume} {93}},\ \bibinfo
  {pages} {184107} (\bibinfo {year} {2016})}\BibitemShut {NoStop}%
\bibitem [{\citenamefont {Shvydko}\ and\ \citenamefont
  {Lindberg}(2012)}]{Shvy12}%
  \BibitemOpen
  \bibfield  {author} {\bibinfo {author} {\bibfnamefont {Y.}~\bibnamefont
  {Shvydko}}\ and\ \bibinfo {author} {\bibfnamefont {R.}~\bibnamefont
  {Lindberg}},\ }\bibfield  {title} {\bibinfo {title} {Spatiotemporal response
  of crystals in x-ray bragg diffraction},\ }\href@noop {} {\bibfield
  {journal} {\bibinfo  {journal} {Phys. Rev. ST Accel. Beams}\ }\textbf
  {\bibinfo {volume} {15}},\ \bibinfo {pages} {100702} (\bibinfo {year}
  {2012})}\BibitemShut {NoStop}%
\bibitem [{\citenamefont {Rodriguez-Fernandez~et al.}(2018)}]{ARF18}%
  \BibitemOpen
  \bibfield  {author} {\bibinfo {author} {\bibfnamefont {A.}~\bibnamefont
  {Rodriguez-Fernandez~et al.}},\ }\bibfield  {title} {\bibinfo {title}
  {Spatial displacement of forward-diffracted x-ray beams by perfect
  crystals},\ }\href@noop {} {\bibfield  {journal} {\bibinfo  {journal} {Acta
  Cryst. A}\ }\textbf {\bibinfo {volume} {74}},\ \bibinfo {pages} {75}
  (\bibinfo {year} {2018})}\BibitemShut {NoStop}%
\bibitem [{\citenamefont {Rodriguez-Fernandez~et al.}(2020)}]{ARF20}%
  \BibitemOpen
  \bibfield  {author} {\bibinfo {author} {\bibfnamefont {A.}~\bibnamefont
  {Rodriguez-Fernandez~et al.}},\ }\bibfield  {title} {\bibinfo {title} {X-ray
  forward diffraction wave-front propagation in \ce{Si} and \ce{C} single
  crystals: simulations and experiments},\ }\href@noop {} {\bibfield  {journal}
  {\bibinfo  {journal} {Proc. of SPIE, Advances in Computational Methods for
  X-Ray Optics V,}\ }\textbf {\bibinfo {volume} {11493}},\ \bibinfo {pages}
  {114930W} (\bibinfo {year} {2020})}\BibitemShut {NoStop}%
\bibitem [{\citenamefont {Tsai}\ \emph {et~al.}(2016)\citenamefont {Tsai},
  \citenamefont {Diaz}, \citenamefont {Menzel},\ and\ \citenamefont
  {Guizar-Sicairos}}]{Tsai26}%
  \BibitemOpen
  \bibfield  {author} {\bibinfo {author} {\bibfnamefont {E.~H.~R.}\
  \bibnamefont {Tsai}}, \bibinfo {author} {\bibfnamefont {A.}~\bibnamefont
  {Diaz}}, \bibinfo {author} {\bibfnamefont {A.}~\bibnamefont {Menzel}},\ and\
  \bibinfo {author} {\bibfnamefont {M.}~\bibnamefont {Guizar-Sicairos}},\
  }\bibfield  {title} {\bibinfo {title} {X-ray ptychography using a distant
  analyzer},\ }\href@noop {} {\bibfield  {journal} {\bibinfo  {journal} {Optics
  Express}\ }\textbf {\bibinfo {volume} {24}},\ \bibinfo {pages} {6441}
  (\bibinfo {year} {2016})}\BibitemShut {NoStop}%
\bibitem [{Note1()}]{Note1}%
  \BibitemOpen
  \bibinfo {note} {We note that conventional ptychography, in which the sample
  is scanned with respect to the incoming beam, would not work in presence of
  dynamical diffraction due the assumption of a factorization of the
  illumination and the sample transmissivity in conventional ptychography, ref.
  \cite {Thibault379}}\BibitemShut {NoStop}%
\bibitem [{\citenamefont {Verezhak~et al.}(2018)}]{Verezhak18}%
  \BibitemOpen
  \bibfield  {author} {\bibinfo {author} {\bibfnamefont {M.}~\bibnamefont
  {Verezhak~et al.}},\ }\bibfield  {title} {\bibinfo {title} {Visualization of
  crystallographic defects in \ce{InSb} micropillars by ptychographic
  topography},\ }\href {https://doi.org/10.1017/S1431927618012527} {\bibfield
  {journal} {\bibinfo  {journal} {Microscopy and Microanalysis}\ }\textbf
  {\bibinfo {volume} {24}},\ \bibinfo {pages} {18} (\bibinfo {year}
  {2018})}\BibitemShut {NoStop}%
\bibitem [{\citenamefont {Verezhak}\ \emph {et~al.}(2021)\citenamefont
  {Verezhak}, \citenamefont {Van~Petegem}, \citenamefont {Rodriguez-Fernandez},
  \citenamefont {Godard}, \citenamefont {Wakonig}, \citenamefont {Karpov},
  \citenamefont {Jacques}, \citenamefont {Menzel}, \citenamefont {Thilly},\
  and\ \citenamefont {Diaz}}]{verezhak2020}%
  \BibitemOpen
  \bibfield  {author} {\bibinfo {author} {\bibfnamefont {M.}~\bibnamefont
  {Verezhak}}, \bibinfo {author} {\bibfnamefont {S.}~\bibnamefont
  {Van~Petegem}}, \bibinfo {author} {\bibfnamefont {A.}~\bibnamefont
  {Rodriguez-Fernandez}}, \bibinfo {author} {\bibfnamefont {P.}~\bibnamefont
  {Godard}}, \bibinfo {author} {\bibfnamefont {K.}~\bibnamefont {Wakonig}},
  \bibinfo {author} {\bibfnamefont {D.}~\bibnamefont {Karpov}}, \bibinfo
  {author} {\bibfnamefont {V.~L.~R.}\ \bibnamefont {Jacques}}, \bibinfo
  {author} {\bibfnamefont {A.}~\bibnamefont {Menzel}}, \bibinfo {author}
  {\bibfnamefont {L.}~\bibnamefont {Thilly}},\ and\ \bibinfo {author}
  {\bibfnamefont {A.}~\bibnamefont {Diaz}},\ }\bibfield  {title} {\bibinfo
  {title} {X-ray ptychographic topography: A robust nondestructive tool for
  strain imaging},\ }\href {https://doi.org/10.1103/PhysRevB.103.144107}
  {\bibfield  {journal} {\bibinfo  {journal} {Phys. Rev. B}\ }\textbf {\bibinfo
  {volume} {103}},\ \bibinfo {pages} {144107} (\bibinfo {year}
  {2021})}\BibitemShut {NoStop}%
\bibitem [{\citenamefont {Batignani~et al.}(2018)}]{Crysta_top}%
  \BibitemOpen
  \bibfield  {author} {\bibinfo {author} {\bibfnamefont {G.}~\bibnamefont
  {Batignani~et al.}},\ }\bibfield  {title} {\bibinfo {title} {Probing
  femtosecond lattice displacement upon photo-carrier generation in lead halide
  perovskite},\ }\href@noop {} {\bibfield  {journal} {\bibinfo  {journal}
  {Nature Communications}\ }\textbf {\bibinfo {volume} {9}},\ \bibinfo {pages}
  {1971} (\bibinfo {year} {2018})}\BibitemShut {NoStop}%
\bibitem [{SM2(2020)}]{SM20}%
  \BibitemOpen
  \href@noop {} {\bibinfo {title} {Supplemental material. \ce{URL} to be
  assigned}} (\bibinfo {year} {2020})\BibitemShut {NoStop}%
\bibitem [{\citenamefont {Reuber}\ \emph {et~al.}(2014)\citenamefont {Reuber},
  \citenamefont {Eisenlohr}, \citenamefont {Roters},\ and\ \citenamefont
  {Raabe}}]{Reuber14}%
  \BibitemOpen
  \bibfield  {author} {\bibinfo {author} {\bibfnamefont {C.}~\bibnamefont
  {Reuber}}, \bibinfo {author} {\bibfnamefont {P.}~\bibnamefont {Eisenlohr}},
  \bibinfo {author} {\bibfnamefont {F.}~\bibnamefont {Roters}},\ and\ \bibinfo
  {author} {\bibfnamefont {D.}~\bibnamefont {Raabe}},\ }\bibfield  {title}
  {\bibinfo {title} {Dislocation density distribution around an indent in
  single-crystalline nickel: Comparing nonlocal crystal plasticity
  finite-element predictions with experiments.},\ }\href@noop {} {\bibfield
  {journal} {\bibinfo  {journal} {Acta Marerialia}\ }\textbf {\bibinfo {volume}
  {71}},\ \bibinfo {pages} {333} (\bibinfo {year} {2014})}\BibitemShut
  {NoStop}%
\bibitem [{\citenamefont {Lie}\ \emph {et~al.}(2014)\citenamefont {Lie},
  \citenamefont {Lu}, \citenamefont {Tieu},\ and\ \citenamefont {Yu}}]{Liu14}%
  \BibitemOpen
  \bibfield  {author} {\bibinfo {author} {\bibfnamefont {M.}~\bibnamefont
  {Lie}}, \bibinfo {author} {\bibfnamefont {C.}~\bibnamefont {Lu}}, \bibinfo
  {author} {\bibfnamefont {K.}~\bibnamefont {Tieu}},\ and\ \bibinfo {author}
  {\bibfnamefont {H.}~\bibnamefont {Yu}},\ }\bibfield  {title} {\bibinfo
  {title} {Numereical comparison between berkovich and conical
  nano-indentations: Mechanical behaviour and micro-texture evolution},\
  }\href@noop {} {\bibfield  {journal} {\bibinfo  {journal} {Materials Science
  {\&} Engineering A}\ }\textbf {\bibinfo {volume} {619}},\ \bibinfo {pages}
  {57} (\bibinfo {year} {2014})}\BibitemShut {NoStop}%
\bibitem [{Note2()}]{Note2}%
  \BibitemOpen
  \bibinfo {note} {From Bragg's law $2d \protect \qopname \relax o{sin}\theta
  =n\lambda $}\BibitemShut {NoStop}%
\bibitem [{\citenamefont {Wakonig~et al.}(2020)}]{Wakonig20}%
  \BibitemOpen
  \bibfield  {author} {\bibinfo {author} {\bibfnamefont {K.}~\bibnamefont
  {Wakonig~et al.}},\ }\bibfield  {title} {\bibinfo {title} {Ptychoshelves, a
  versatile highlevel framework for high-performance analysis of ptychographic
  data},\ }\href@noop {} {\bibfield  {journal} {\bibinfo  {journal} {Journal of
  applied crystallography}\ }\textbf {\bibinfo {volume} {53}},\ \bibinfo
  {pages} {574} (\bibinfo {year} {2020})}\BibitemShut {NoStop}%
\bibitem [{\citenamefont {Thibault~et al.}(2009)}]{Thibault09}%
  \BibitemOpen
  \bibfield  {author} {\bibinfo {author} {\bibfnamefont {P.}~\bibnamefont
  {Thibault~et al.}},\ }\bibfield  {title} {\bibinfo {title} {Probe retrieval
  in ptychographic coherent diffractive imaging},\ }\href@noop {} {\bibfield
  {journal} {\bibinfo  {journal} {Ultramicroscopy}\ }\textbf {\bibinfo {volume}
  {109}},\ \bibinfo {pages} {338} (\bibinfo {year} {2009})}\BibitemShut
  {NoStop}%
\bibitem [{\citenamefont {Thibault}\ and\ \citenamefont
  {Guizar-Sicairos}(2012)}]{Thibault12}%
  \BibitemOpen
  \bibfield  {author} {\bibinfo {author} {\bibfnamefont {P.}~\bibnamefont
  {Thibault}}\ and\ \bibinfo {author} {\bibfnamefont {M.}~\bibnamefont
  {Guizar-Sicairos}},\ }\bibfield  {title} {\bibinfo {title}
  {Maximum-likelihood refinement for coherent diffractive imaging},\
  }\href@noop {} {\bibfield  {journal} {\bibinfo  {journal} {New Journal of
  Physics}\ }\textbf {\bibinfo {volume} {14}} (\bibinfo {year}
  {2012})}\BibitemShut {NoStop}%
\bibitem [{\citenamefont {Guizar-Sicairos~et al.}(2014)}]{Guizar14}%
  \BibitemOpen
  \bibfield  {author} {\bibinfo {author} {\bibfnamefont {M.}~\bibnamefont
  {Guizar-Sicairos~et al.}},\ }\bibfield  {title} {\bibinfo {title}
  {High-throughput ptychography using eiger: scanning x-ray nano-imaging of
  extended regions},\ }\href@noop {} {\bibfield  {journal} {\bibinfo  {journal}
  {Optics Express}\ }\textbf {\bibinfo {volume} {22}},\ \bibinfo {pages}
  {14859} (\bibinfo {year} {2014})}\BibitemShut {NoStop}%
\bibitem [{\citenamefont {van Heel}\ and\ \citenamefont
  {Schatz}(2005)}]{Heel05}%
  \BibitemOpen
  \bibfield  {author} {\bibinfo {author} {\bibfnamefont {M.}~\bibnamefont {van
  Heel}}\ and\ \bibinfo {author} {\bibfnamefont {M.}~\bibnamefont {Schatz}},\
  }\bibfield  {title} {\bibinfo {title} {Fourier shell correlation threshold
  criteria},\ }\href@noop {} {\bibfield  {journal} {\bibinfo  {journal}
  {Journal of Structural Biology}\ }\textbf {\bibinfo {volume} {151}},\
  \bibinfo {pages} {250} (\bibinfo {year} {2005})}\BibitemShut {NoStop}%
\bibitem [{\citenamefont {Lings~et al.}(2006)}]{Lings06}%
  \BibitemOpen
  \bibfield  {author} {\bibinfo {author} {\bibfnamefont {B.}~\bibnamefont
  {Lings~et al.}},\ }\bibfield  {title} {\bibinfo {title} {Simulations of
  time-resolved x-ray diffraction in laue geometry},\ }\href@noop {} {\bibfield
   {journal} {\bibinfo  {journal} {J. Phys.: Condens. Matter}\ }\textbf
  {\bibinfo {volume} {18}},\ \bibinfo {pages} {9231–9244} (\bibinfo {year}
  {2006})}\BibitemShut {NoStop}%
\bibitem [{\citenamefont {Takagi}(1962)}]{Takagi62}%
  \BibitemOpen
  \bibfield  {author} {\bibinfo {author} {\bibfnamefont {S.}~\bibnamefont
  {Takagi}},\ }\bibfield  {title} {\bibinfo {title} {Dynamical theory of
  diffaction applicable to cystals with any kind of small distortions},\
  }\href@noop {} {\bibfield  {journal} {\bibinfo  {journal} {Acta. Cryst.}\
  }\textbf {\bibinfo {volume} {15}},\ \bibinfo {pages} {1311} (\bibinfo {year}
  {1962})}\BibitemShut {NoStop}%
\bibitem [{\citenamefont {Kato}\ and\ \citenamefont {Lang}(1959)}]{Kato59}%
  \BibitemOpen
  \bibfield  {author} {\bibinfo {author} {\bibfnamefont {N.}~\bibnamefont
  {Kato}}\ and\ \bibinfo {author} {\bibfnamefont {A.~R.}\ \bibnamefont
  {Lang}},\ }\bibfield  {title} {\bibinfo {title} {A study of
  pendell{\"{o}}sung fringes in x-ray diffraction},\ }\href@noop {} {\bibfield
  {journal} {\bibinfo  {journal} {Acta Cryst.}\ }\textbf {\bibinfo {volume}
  {12}},\ \bibinfo {pages} {787} (\bibinfo {year} {1959})}\BibitemShut
  {NoStop}%
\bibitem [{\citenamefont {Li~et al}(2015)}]{Li2015}%
  \BibitemOpen
  \bibfield  {author} {\bibinfo {author} {\bibfnamefont {Z.~J.}\ \bibnamefont
  {Li~et al}},\ }\bibfield  {title} {\bibinfo {title} {{Local strain and
  defects in silicon wafers due to nanoindentation revealed by full-field X-ray
  microdiffraction imaging}},\ }\href
  {https://doi.org/10.1107/S1600577515009650} {\bibfield  {journal} {\bibinfo
  {journal} {Journal of Synchrotron Radiation}\ }\textbf {\bibinfo {volume}
  {22}},\ \bibinfo {pages} {1083} (\bibinfo {year} {2015})}\BibitemShut
  {NoStop}%
\end{thebibliography}%
 
\end{document}